\documentclass[preprint]{revtex4}
\usepackage{txfonts}

\usepackage{graphicx}% Include figure files]
\usepackage{bm}% bold math
\usepackage{color}

\begin{document}
\title{{Magnetization process of spin-1/2 Heisenberg antiferromagnets on a layered triangular lattice}}

\author{Daisuke Yamamoto$^{1}$}
\author{Giacomo Marmorini$^{2,3}$}
\author{Ippei Danshita$^{2}$}
\affiliation{
{$^1$Waseda Institute for Advanced Study, Waseda University, Tokyo 169-8050, Japan}
\\
{$^2$Yukawa Institute for Theoretical Physics, Kyoto University, Kyoto 606-8502, Japan}
\\
{$^3$Research and Education Center for Natural Sciences, Keio University, Kanagawa 223-8521, Japan}
}

\begin{abstract}
We study the magnetization process of the spin-1/2 antiferromagnetic Heisenberg model on a layered triangular lattice by means of a numerical cluster mean-field method with a scaling scheme (CMF+S). It has been known that antiferromagnetic spins on a two-dimensional (2D) triangular lattice with quantum fluctuations exhibit a one-third magnetization plateau in the magnetization curve under magnetic field. We demonstrate that the CMF+S quantitatively reproduces the magnetization curve including the stabilization of the plateau. {We also discuss the effects of a finite interlayer coupling, which is unavoidable in real quasi-2D materials. It has been recently argued for a model of the layered-triangular-lattice compound Ba$_3$CoSb$_2$O$_9$ that such interlayer coupling can induce an additional first-order transition at a strong field. We present the detailed CMF+S results for the magnetization and susceptibility curves of the fundamental Heisenberg Hamiltonian in the presence of magnetic field and weak antiferromagnetic interlayer coupling. The extra first-order transition appears as a quite small jump in the magnetization curve and a divergence in the susceptibility at a strong magnetic field $\sim 0.712$ of the saturation field. } 
\end{abstract}
%%% Keywords are not needed any longer. %%%
%%%\kword{keyword1, keyword2, keyword3, \ldots}
%%%

\maketitle

\section{Introduction}
\label{sect:introduction}

Triangular-lattice antiferromagnets (TLAFs), which are a paradigmatic example of geometric frustration, have received renewed interest~\cite{starykh-15} in recent years owing to the technical developments in high-field experiments and the appearance of new materials comprising Co$^{2+}$ magnetic ions, such as Ba$_3$CoSb$_2$O$_9$~\cite{shirata-12,zhou-12,susuki-13,koutroulakis-14,quirion-15,ma-15} and Ba$_3$CoNb$_2$O$_9$~\cite{lee-14,yokota-14,sun-15}. Unlike other typical TLAF compounds with Cu$^{2+}$ ions, such as Cs$_2$CuCl$_4$ and Cs$_2$CuBr$_4$~\cite{ono-03,fortune-09}, the Co-based compounds can form regular (undistorted) triangular-lattice layers and are free from a large antisymmetric interaction of the Dzyaloshinsky-Moriya type thanks to the highly symmetric crystal structure. The physics of those compounds {is} expected to be described by a simple model Hamiltonian.

Recently, Shirata $et$ $al$.~\cite{shirata-12} reported that the magnetization curve of Ba$_3$CoSb$_2$O$_9$ powder seems to show excellent agreement with theoretical calculations on the spin-1/2 isotropic Heisenberg model on a two-dimensional (2D) triangular lattice~\cite{chubukov-91,nishimori-86,honecker-04,yoshikawa-04,farnell-09,sakai-11,gotzea-16}. This is owing to the fact that the magnetic layers of Co$^{2+}$ ions are well separated from each other by a nonmagnetic layer~\cite{shirata-12}. However, the latest experiments with the use of single crystals~\cite{zhou-12,susuki-13,koutroulakis-14,quirion-15} found a field-direction dependence of magnetization curve, which indicates the existence of exchange anisotropy, and a magnetization anomaly that had not been predicted at a strong magnetic field $\sim 22$ T perpendicular to the $c$ axis. Several conjectures have been proposed for the origin of the unexpected high-field anomaly in the magnetization curve~\cite{susuki-13,koutroulakis-14,maryasin-13,yamamoto-15,sellmann-15}.

In our previous study~\cite{yamamoto-15}, we have provided a microscopic model calculation for the magnetization process of the quasi-2D TLAF Ba$_3$CoSb$_2$O$_9$ by taking into account the easy-plane exchange anisotropy and weak couplings between layers. The theoretical magnetization curve under in-plane magnetic field exhibits a field-induced first-order transition at $\sim 0.7$ of the saturation field $H_s$ as well as the well-known plateau structure at the one-third of the saturation magnetization. From the result, we suggested that the origin of the magnetization anomaly observed in Ba$_3$CoSb$_2$O$_9$ is indeed the extra first-order transition due to the weak interlayer coupling. In fact, the critical field strength ($\sim 22$ T) at the anomaly is well accorded with our theoretical prediction ($\sim 0.7H_s$~\cite{yamamoto-15}), {given that} the saturation field of Ba$_3$CoSb$_2$O$_9$ is about $31.9$ T~\cite{susuki-13}.

In Ref.~\onlinecite{yamamoto-15}, we focused on the system with easy-plane exchange anisotropy in order to explain a specific experiment. In this paper, we give a quantitative study on quantum TLAFs based on the isotropic Heisenberg Hamiltonian, which is a simpler but more fundamental model to describe spin systems. Even for the simple Heisenberg model on a single layer of triangular lattice, there have been only a few numerical studies~\cite{nishimori-86,honecker-04,yoshikawa-04,farnell-09,sakai-11,gotzea-16} that can reproduce the quantum stabilization of the magnetization plateau. This is due to the difficulty in treating strongly frustrated quantum systems. Also in experiments, there are only a few quantum TLAF materials that actually exhibit a clear magnetization plateau~\cite{ono-03,fortune-09,shirata-12}. Therefore, it is still important to reproduce the plateau of the isotropic TLAF by different theoretical methods, {whose reliability can in turn be confirmed by their accurate  prediction of the plateau.}

{We apply our numerical cluster mean-field approach with a scaling scheme (CMF+S)~\cite{yamamoto-12-1,yamamoto-12-2,yamamoto-13-2,yamamoto-14,morenocardone-14} to the $S=1/2$ Heisenberg model on a triangular lattice. The result for a single layer shows that the numerical CMF+S {method} provides an excellent quantitative agreement with previous numerical results~\cite{chubukov-91,honecker-04,farnell-09,sakai-11,gotzea-16}, including the quantum stabilization of the one-third magnetization plateau. We also present the CMF+S results for the magnetization and susceptibility curves in the presence of antiferromagnetic interlayer coupling. It is shown that a small interlayer coupling gives rise to an extra discontinuous quantum phase transition at a strong field $\approx 0.712H_s$. The high-field first-order transition is clearly visible as a divergence in the susceptibility while the shape of the magnetization curve undergoes very little change except for {quite a} small jump at the transition point. }

\section{The spin-1/2 Heisenberg Model on a Layered Triangular Lattice}
\begin{figure}[t]
\begin{center}
\includegraphics[scale=0.15]{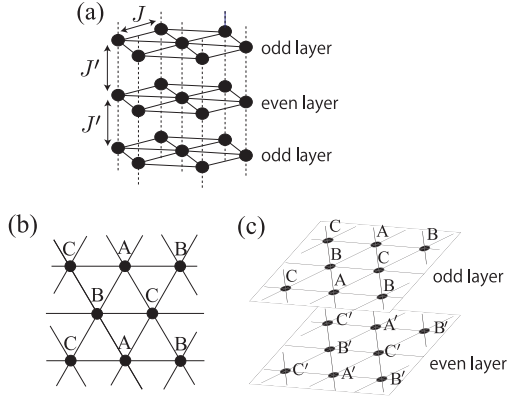}
\caption{\label{figTLAF}
(a) Weakly-coupled layers of triangular lattice. (b) A three-sublattice spin structure expected on a triangular lattice. The spins on the sites with the same letter points in the same direction. (c) A six-sublattice spin structure on a layered triangular lattice. }
\end{center}
\end{figure}
The Hamiltonian of the Heisenberg model on layers of triangular lattice [see Fig.~\ref{figTLAF}(a)] is given by
\begin{eqnarray}
\hat{\mathcal{H}}=J\sum_{\langle i,j\rangle}\hat{\bm{S}}_i\cdot\hat{\bm{S}}_j+J^\prime\sum_{\langle i,l\rangle^\prime}\hat{\bm{S}}_i\cdot\hat{\bm{S}}_l-H\sum_{i}\hat{S}^z_{i},
\label{hamiltonian}
\end{eqnarray}
where both the intralayer ($J$) and interlayer ($J^\prime$) nearest-neighbor (NN) couplings are assumed to be antiferromagnetic. We investigate the $S=1/2$ case that exhibits the strongest quantum fluctuations.

For a single layer of triangular lattice (or $J^\prime=0$), the ground-state spin configuration usually forms a three-sublattice structure {as} shown in Fig.~\ref{figTLAF}(b) as long as the system only has spatially isotropic NN interactions. However, real TLAF compounds such as Ba$_3$CoSb$_2$O$_9$ have a quasi-2D structure that consists of magnetic triangular-lattice layers separated from each other by a nonmagnetic layer~\cite{shirata-12}. Therefore, one has to consider a small but finite interlayer interaction $J^\prime$ between spins on different layers. For a ferromagnetic $J^\prime<0$, it is expected that the spins along the stacking direction tend to point in the same direction and the three-sublattice spin configuration on each layer is more stabilized. A quite small $|J^\prime|\ll J$ can be thus negligible when the magnetic layers are well separated. However, we will show that when $J^\prime>0$, even a small interlayer interaction could play an essential role in determining the ground-state spin configuration of TLAFs under magnetic field. The consideration of this effect due to weak three dimensionality is a key to quantitatively describe quasi-2D quantum TLAF compounds with a microscopic model Hamiltonian.

\section{Cluster Mean-Field Approach with Cluster-Size Scaling for Quantum Spins}
\label{sect:CMF+S}

The classical counterpart of the model ($\ref{hamiltonian}$) is written in terms of three-dimensional vector $\bm{S}_i$ with a fixed length $|\bm{S}_i|=1/2$ instead of quantum spin-1/2 operator $\hat{\bm{S}}_i$. It is known that the ground state of the classical Heisenberg model on a 2D triangular lattice ($J^\prime=0$) cannot be uniquely determined under magnetic field ($H\neq 0$)~\cite{kawamura-85,capriotti-98}. Although the classical-spin angles on the three-sublattice structure shown in Fig.~\ref{figTLAF}(a) have six degrees of freedom, the minimization of the classical energy only gives the constraint $\bm{S}_{ A}+\bm{S}_{ B}+\bm{S}_{ C}=(0,0,H/3J)$ for $0<H<H_s$~\cite{kawamura-85,capriotti-98}. Here, $H_s=9J/2$ is the saturation field, above which all the spins are aligned parallel to the magnetic field. Except the trivial degeneracy from the $U(1)$ symmetry, there still remain two degrees of freedom. Therefore, spin configurations in the ground-state magnetization process ($0<H<H_s$) are continuously degenerate. The total magnetization per site, $M\equiv \sum_i S_i^z/N$ with $N$ being the number of lattice sites, is given by $H/9J$ as a linear function of the magnetic field strength for any magnetization process in the degenerate manifold.

\begin{figure*}[t]
\begin{center}
\includegraphics[scale=0.4]{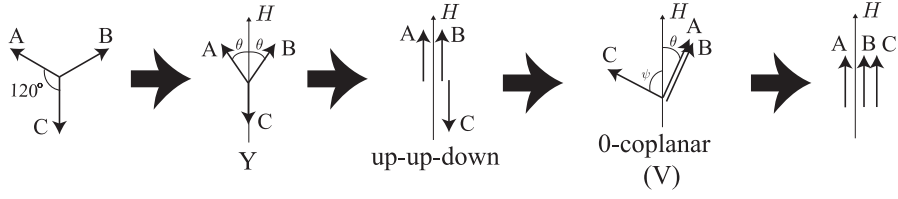}
\caption{\label{fig2D}
The ground-state magnetization process of quantum Heisenberg antiferromagnets on a purely 2D triangular lattice. Each arrow represents the spin angle on the sublattice A, B, or C. The 0-coplanar state is also called the ``V'' state.}
\end{center}
\end{figure*}

In the quantum spin-1/2 system, however, the classical degeneracy is lifted by quantum fluctuation effects. For TLAFs, it is now established that the sequence of the ``Y,'' up-up-down, and 0-coplanar states [Fig.~\ref{fig2D}] is selected as the lowest-energy magnetization process~\cite{chubukov-91,nishimori-86,honecker-04,yoshikawa-04,farnell-09,sakai-11}. The magnetization curve $M(H)$ is no longer a linear function of $H$, and exhibits a magnetization plateau at one-third of the saturation magnetization reflecting the stabilization of the up-up-down state over a finite range of magnetic field strength.
\begin{figure}[t]
\begin{center}
\includegraphics[scale=0.4]{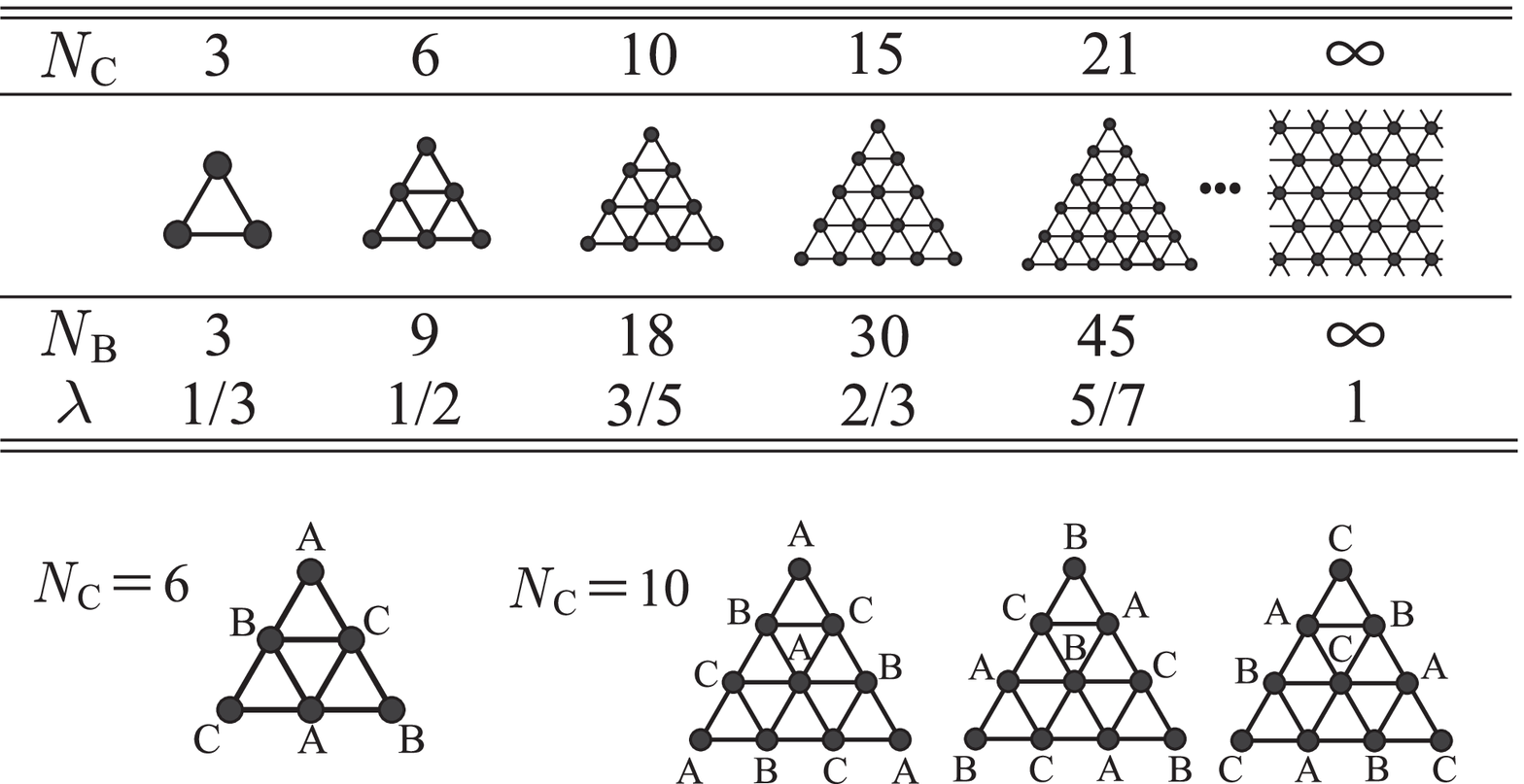}
\caption{\label{figclusters}
Series of clusters used in the present CMF+S study. The lower illustrations show the independent clusters that have to be considered under the three-sublattice ansatz in the purely 2D case. The cluster of $N_C=3$, 15, or 21 is similar to that of $N_C=6$. 
}
\end{center}
\end{figure}
We employ here the CMF+S method~\cite{yamamoto-14,yamamoto-12-1,yamamoto-12-2,yamamoto-13-2,morenocardone-14} to {take into account} the quantum effects and interlayer coupling. In general, the number of spins that can be numerically diagonalized on current computer resources is very limited. To avoid strong finite-size effects in such exact diagonalization on small clusters, we impose a self-consistent mean-field boundary condition instead of the usual periodic or open one. First, we approximate the Hamiltonian $\hat{\mathcal{H}}$ on $N$ sites ($N\rightarrow \infty$ in the thermodynamic limit) by the sum of $N/N_C$ cluster Hamiltonians on a cluster of $N_C$ sites. The inter-cluster interactions are decoupled as $\hat{S}_{i}^{\alpha} \hat{S}_{j}^\alpha \rightarrow \langle \hat{S}_{i}^{\alpha} \rangle \hat{S}_{j}^\alpha + \langle \hat{S}_{j}^{\alpha} \rangle \hat{S}_{i}^\alpha$ ($\alpha=\{x,y,z\}$). Thus, the cluster Hamiltonian $\hat{\mathcal{H}}_{C_n}$ includes the expectation values $\langle \hat{S}_{i}^{\alpha} \rangle$ as mean fields to be determined self-consistently. The sublattice magnetic moment $m^\alpha_\mu$ on sublattice $\mu$ is given by
\begin{eqnarray}
m^\alpha_\mu=  \frac{1}{N_\mu}\sum_{n}\sum_{i_\mu\in C_n}{\rm Tr}\left( \hat{S}_{i_\mu}^{\alpha}e^{-\beta \hat{\mathcal{H}}_{C_n}}\right)\Big{/} {\rm Tr}( e^{-\beta \hat{\mathcal{H}}_{C_n}}), \label{selfconsistent}
\end{eqnarray}
where $n=1,2,\cdots,M_C$ with $M_C$ being the number of the independent clusters (specifically given later), $N_\mu$ is the number of total sites belonging to the sublattice $\mu$ in the $M_C$ clusters, and $\beta=1/T$ (we take $T\rightarrow 0$ in the present study). Substituting $m^\alpha_\mu$ into $\langle \hat{S}_{i_\mu}^{\alpha} \rangle$ in $\hat{\mathcal{H}}_{C_n}$, Eq.~(\ref{selfconsistent}) becomes a set of self-consistent equations for $m^\alpha_\mu$.

Usually, one gets some different sets of solutions for $m^\alpha_\mu$ depending on the initial values in the iteration process to solve the self-consistent equations. The spin configuration of each solution at a given magnetic field strength $H$ is identified by the converged values of $m^\alpha_\mu$ on each sublattice $\mu$. Since the energy difference between different spin configurations can be estimated by integrating the magnetization curve $M(H)$ (which is given by the average of $m_\mu^z$ over $\mu$) with respect to $H$ from $H_s$, one can determine {the ground-state magnetization process} by comparing the energies of the different solutions.

To efficiently treat possible spin configurations in quasi-2D TLAFs, we choose a series of triangular-shaped clusters of up to $N_C=21$, displayed in Fig.~\ref{figclusters}. The above-mentioned approach reproduces the classical ground state for $N_C=1$, and allows for a systematic inclusion of non-local fluctuations as $N_C$ increases. Therefore, we eventually make an extrapolation of the results for different values of $N_C$ to the limit of $N_C \rightarrow \infty$, where long-range fluctuations in each triangular-lattice layer are fully included. The scaling parameter $\lambda \equiv N_B/3N_C$ ($N_B$ is the number of bonds treated exactly) varies from 0 for $N_C=1$ to 1 for $N_C=\infty$.

\section{Purely Two-Dimensional Case ($J^\prime=0$)}
\label{sect:2D}

We first present the CMF+S result for the magnetization curve of the Heisenberg model on a single layer of triangular lattice ($J^\prime=0$) for comparisons with some known results obtained by other theoretical methods~\cite{chubukov-91,honecker-04,farnell-09,sakai-11}. Under the assumption of the three-sublattice structure shown in Fig.~\ref{figTLAF}(b), the number of independent clusters that have to be treated in the CMF+S self-consistent equations (\ref{selfconsistent}) is $M_C=1$ for $N_C=3$, 6, 15, or 21, while $M_C=3$ for $N_C=10$ (see the lower illustrations of Fig.~\ref{figclusters}).

\begin{figure*}[t]
\begin{center}
\includegraphics[scale=0.3]{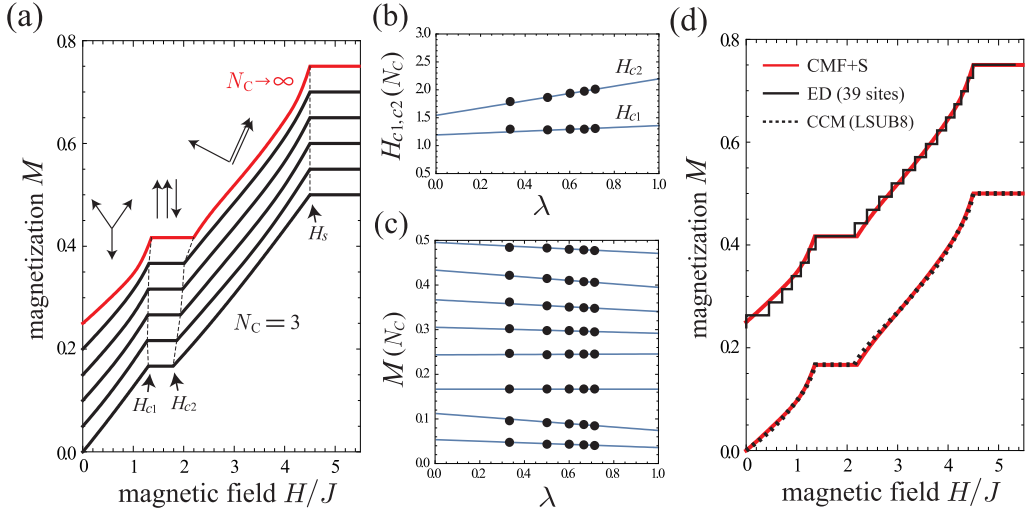}
\caption{\label{figMagnetization2D}
(a) Ground-state magnetization curves of the Heisenberg model on a single layer of triangular lattice for $N_C=3$, 6, 10, 15, 21, and $\infty$ (from bottom to top). All the curves apart from the bottom one are vertically shifted by 0.05, 0.1, 0.15, 0.2, 0.25, respectively, for clarity. (b) Cluster-size scalings of the data for the phase transition points $H_{c1}$ and $H_{c2}$. (c) Cluster-size scalings of the data for the magnetization $M$ at the rescaled field strength $H^\prime=0.4$ and 0.8 (lower two lines), at $H_{c1}\leq H^\prime\leq H_{c2}$ (the line of $M=1/6$), and at $H^\prime=2.8$, 3.2, 3.6, 4.0, and 4.4 (upper five lines). (d) Comparisons with the exact diagonalization (ED) with the periodic boundary condition~\cite{sakai-11} and with the coupled cluster method (CCM)~\cite{farnell-09}. The upper curves are vertically shifted by 0.25 for clarity. }
\end{center}
\end{figure*}

The obtained magnetization curves $M(H)$ for each $N_C$ are shown in Fig.~\ref{figMagnetization2D}(a). The quantum fluctuations select the sequence of the Y, up-up-down, and 0-coplanar states as expected. {The} magnetization curve exhibits the one-third quantum magnetization plateau. {As the size of the cluster $N_C$ increases, the plateau gets wider, indicating that the effects of quantum fluctuations are properly included.} Figure~\ref{figMagnetization2D}(b) shows the critical field strength at each of end points of the plateau (named $H_{c1}$ and $H_{c2}$) as a function of the scaling parameter $\lambda$. We perform a linear extrapolation $N_C\rightarrow \infty$ ($\lambda\rightarrow 1$) of the data calculated with the three largest clusters. The extrapolated values are given as $H_{c1}(\infty)/J=1.359$ and $H_{c2}(\infty)/J=2.196$, respectively. These values are {somewhat different from the linear-spin-wave result (1.248 and 2.145~\cite{chubukov-91}) and} comparable with those obtained by the exact diagonalization with the periodic boundary condition (1.38 and 2.16~\cite{honecker-04,sakai-11}) and by the coupled cluster method (1.37 and 2.15~\cite{farnell-09}).

 {In the previous CMF+S study on the $XXZ$ Hamiltonian~\cite{yamamoto-14} we derived the phase diagram in the plane of the $XXZ$ anisotropy and magnetic field strength by extrapolating the values of the anisotropy parameter {$J/J_z$} {at each phase boundary with $H/H_s$ fixed. Since we took sufficiently many but necessarily a finite number of $H/H_s$ values, a further interpolation was needed to obtain the critical fields corresponding to the isotropic Heisenberg model, namely to the line $J/J_z=1$. This procedure gave $H_{c1}/J=1.345$ and $H_{c2}/J=2.113$~\cite{yamamoto-14}. It is expected that the values obtained by the present work ($1.359$ and $2.196$ {respectively}) are more accurate for the following two reasons. First, the more direct scaling procedure for the magnetization curve of the Heisenberg model does not require further interpolations.} {Second,} {the extrapolation to the infinite-size limit based on the linear fit appears to work even more efficiently for the present treatment (in which the critical $H/H_s$ is extrapolated) compared to the $XXZ$ model study (in which the critical $J/J_z$ is extrapolated). }}

In order to make cluster-size scaling of the entire magnetization curve $M(H)$, first we have to change the scale of each curve obtained with finite $N_C$ with respect to $H$ {as $H\rightarrow H^\prime$ with 
\begin{eqnarray*}
H^\prime&=& \frac{H_{c1}(\infty)}{H_{c1}(N_C)}H~~{\rm for}~~0\leq H\leq H_{c1}(N_C),\\
H^\prime&=&\frac{H_{c2}(\infty)-H_{c1}(\infty)}{H_{c2}(N_C)-H_{c1}(N_C)}\big(H-H_{c1}(N_C)\big)+H_{c1}(\infty)\\&&~~{\rm for}~~H_{c1}(N_C)\leq H\leq H_{c2}(N_C),
\end{eqnarray*}
and
\begin{eqnarray*}
H^\prime&=& \frac{H_{s}-H_{c2}(\infty)}{H_{s}-H_{c2}(N_C)}\big(H-H_{c2}(N_C)\big)+H_{c2}(\infty)\\&&~~{\rm for}~~H_{c2}(N_C)\leq H\leq H_{s}
\end{eqnarray*}
}so that the phase boundaries for all finite $N_C$ {have the same location as} those of the limit $N_C\rightarrow \infty$ ($H_{c1}(\infty)$ and $H_{c2}(\infty)$)~\cite{yamamoto-15}. 
The extrapolation of the magnetization $M$ at different values of the rescaled $H^\prime$ are shown in Fig.~\ref{figMagnetization2D}(c). %Note that the magnetization takes $M=1/6$ in the up-up-down state and $M=1/2$ in the saturated state independently of $N_C$. 
The top curve in Fig.~\ref{figMagnetization2D}(a) is the obtained CMF+S magnetization curve, which is compared with the other numerical results~\cite{sakai-11,farnell-09} in Fig.~\ref{figMagnetization2D}(d). We can see an excellent agreement of the CMF+S curve with the data extracted from Refs.~\onlinecite{sakai-11} and~\onlinecite{farnell-09}, regarding the locations of the phase transition points and the nonlinear bending due to the quantum effects.

\section{The Role of Weak Interlayer Coupling ($J^\prime\neq 0$)}
\label{sect:3D}

{Weak three-dimensionality is not avoidable in real quasi-2D TLAF compounds.} Since there are numerous nearly-degenerate states in frustrated systems, even very small interlayer interaction can compete with quantum fluctuations and affect the ground-state selection. {In the previous study~\cite{yamamoto-15}, while {effects of the interlayer coupling have} been discussed with a focus on the case where easy-plane exchange anisotropy exists, we have not explained the detailed magnetization process for the isotropic case. Therefore, here let us discuss the role of interlayer coupling $J^\prime$ on the magnetization process of the fundamental Heisenberg model (\ref{hamiltonian}) in a quantitative way with the use of the CMF+S.}

For weakly-coupled triangular-lattice layers, we assume the six sublattice structure shown in Fig.~\ref{figTLAF}(c) ($\mu=$A, B, C, A$^\prime$, B$^\prime$, or C$^\prime$)~\cite{susuki-13}. The number of independent clusters that have to be considered in the CMF+S is now $M_C=2$ for $N_C=3$, 6, 15, or 21, while $M_C=6$ for $N_C=10$. The saturation field is given by $H_{s}=9J/2+2J^\prime$. 

\begin{figure}[b]
\begin{center}
\includegraphics[scale=0.45]{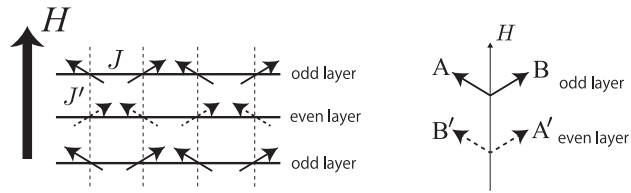}
\caption{\label{fig3D}
Typical spin stacking pattern in weakly-coupled antiferromagnetic layers of bipartite lattice under magnetic field. The right panel is the corresponding schematic illustration. 
}
\end{center}
\end{figure}

{In general, when} the interlayer interaction is antiferromagnetic ($J^\prime>0$), antiparallel spin alignment is favored along the stacking direction. If each layer consists of a bipartite lattice, it is expected that the in-plane magnetic order is just stacked alternately as shown in Fig.~\ref{fig3D}. However, for non-bipartite lattices such as the triangular lattice, the in-plane three-sublattice magnetic order could compete with the demand of antiparallel alignment along the stacking direction. As a result of the incompatibility, the lowest-energy stacking pattern can be changed depending on the magnetic field strength as will be shown below.

\begin{figure*}[t]
\begin{center}
\includegraphics[scale=0.35]{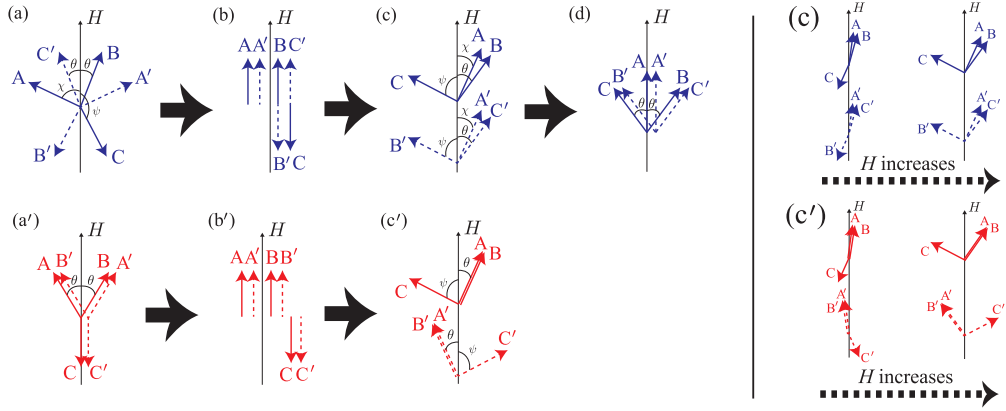}
\caption{\label{figbranch}
Two candidate magnetization processes of quasi-2D Heisenberg TLAFs. The interlayer interaction $J^\prime$ acts on the NN bonds between A and A$^\prime$, between B and B$^\prime$, and between C and C$^\prime$. The right panels show the changes of relative angles $\theta$, $\phi$, and $\chi$ in the states (c) and (c$^\prime$) as $H$ increases. 
}
\end{center}
\end{figure*}
For {TLAFs in the presence of} antiferromagnetic interlayer coupling, we found two candidate magnetization processes as solutions of Eq.~(\ref{selfconsistent})~\cite{yamamoto-15}, both of which reduce to the sequence of the Y, up-up-down, and 0-coplanar states at the purely 2D limit. Figure~\ref{figbranch} shows the two branches (a)-(b)-(c)-(d) and (a$^\prime$)-(b$^\prime$)-(c$^\prime$). From comparison between (a) and (a$^\prime$) or between (b) and (b$^\prime$), it is obviously seen that the former process has a lower energy for low magnetic fields since $J^\prime>0$ favors antiparallel alignment on interlayer bonds. However, the magnitude relation between the energies of (c) and (c$^\prime$) at strong fields can be reversed as $H$ increases. Note that the relative angles ($\theta$, $\phi$, and $\chi$) among the six sublattice magnetic moments are gradually changed as a function of $H$ even in the same phase. As can be seen in the right panels of Fig.~\ref{figbranch}, when the spins are almost collinear, the energy of (c$^\prime$) should be higher than that of (c) due to the high interlayer bond energy of field-parallel spin components. However, when the magnetic field is further {increased}, the (c$^\prime$) configuration becomes advantageous against (c) since it can reduce more interlayer bond energy of field-transverse components. As a result, a field-induced first-order transition between (c) and (c$^\prime$) is expected to occur at a certain strong magnetic field. 
\begin{figure*}[t]
\begin{center}
\includegraphics[scale=0.3]{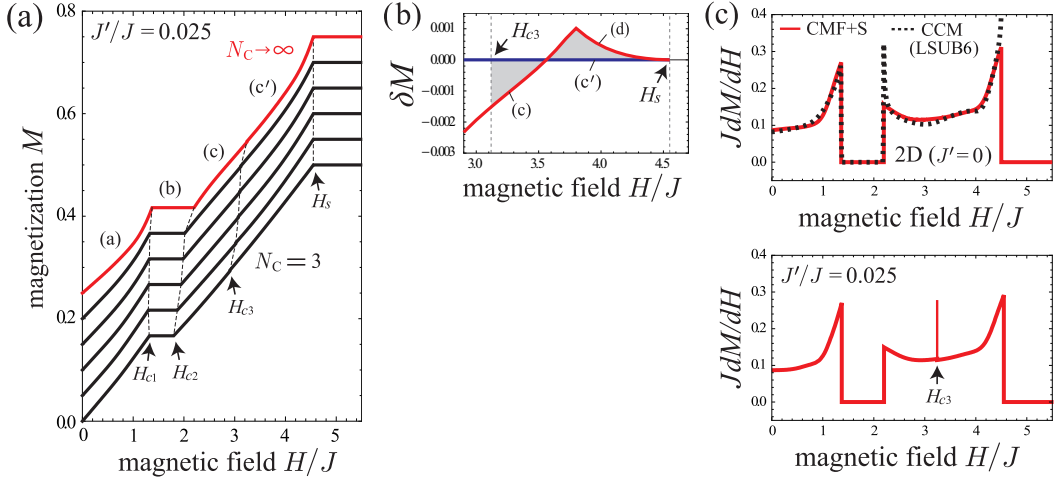}
\caption{\label{figMagnetization3D}
(a) Ground-state magnetization curves of the Heisenberg model on weakly-coupled layers of triangular lattice for $N_C=3$, 6, 10, 15, 21, and $\infty$ (from bottom to top). We set $J^\prime=0.025J$. All the curves apart from the bottom one are vertically shifted by 0.05, 0.1, 0.15, 0.2, 0.25, respectively, for clarity. (b) The high-field part of the magnetization curves of the two branches. Here, $\delta M$ is the magnetization measured from that of the branch (a$^\prime$)-(b$^\prime$)-(c$^\prime$). The critical field strength $H_{c3}$ is determined so that the areas of the shaded regions are equal. The result for $N_C=21$ is shown. (c) The CMF+S result ($N_C\rightarrow \infty$) of the susceptibility $J dM/dH$ for $J^\prime=0$ (upper) and $J^\prime=0.025J$ (lower). The susceptibility {of} a single layer of triangular lattice ($J^\prime=0$) is compared with that of the CCM~\cite{farnell-09}. 
}
\end{center}
\end{figure*}

Figure~\ref{figMagnetization3D}(a) shows the ground-state magnetization process obtained by solving Eq.~(\ref{selfconsistent}) for different $N_C$ in the case of a very small interlayer interaction $J^\prime=0.025 J$. Indeed, an extra discontinuous phase transition that was not seen in the purely 2D case is found at a strong field, although the discontinuity in $M$ is quite small. The transition point $H_{c3}$ can be determined by Maxwell's construction for the difference of the curves $M(H)$ in the two magnetization processes [see Fig.~\ref{figMagnetization3D}(b)]. Thus, it is concluded that the magnetization process is given by (a)-(b)-(c)-(c$^\prime$) for quasi-2D TLAFs.

Performing a cluster-size extrapolation ($N_C\rightarrow \infty$) in a similar way to the purely 2D case, we obtain the CMF+S result of the magnetization curve [the top curve in Fig.~\ref{figMagnetization3D}(a)]. Since $J^\prime$ is assumed here to be very small, the general behavior of the curve differs little from that of the purely 2D case. However, as shown in Fig.~\ref{figMagnetization3D}(c), the existence of the additional quantum first-order transition is clearly seen in the field derivative of $M(H)$ as the divergence at {$H=H_{c3}\approx 0.712H_s$}, in contrast to the purely 2D ($J^\prime=0$) case.

%------------------------------------------------------------------------------
\section{Summary}
\label{sect:Summary}
In summary, we have studied the magnetization process of the spin-1/2 Heisenberg model on layered triangular lattice with and without weak interlayer coupling. Our numerical calculations with the CMF+S method properly described the one-third magnetization plateau expected in quantum TLAFs and provided a quantitative agreement with the numerical data of the ED~\cite{honecker-04,sakai-11} and the CCM~\cite{farnell-09} for a single layer of triangular lattice. 
{Moreover, we discussed the detailed magnetization process in the presence of weak interlayer coupling, which is unavoidable in real quasi-2D compounds. We presented the magnetization and susceptibility curves of the isotropic Heisenberg model, which {was not reported} in our previous study~\cite{yamamoto-15}. Although a small interlayer coupling does not change the apparent shape of the magnetization curve, an additional first-order phase transition occurs at $H\approx 0.712H_s$ above the one-third plateau. This extra transition is visible as a small discontinuous jump in the magnetization curve and a divergence in the susceptibility. From a comparison of the present isotropic Heisenberg model with the previous easy-plane case~\cite{yamamoto-15}, the easy-plane anisotropy seems not to be relevant for the qualitative feature of the magnetization process and the shape of the magnetization curve including the extra first-order transition as long as the magnetic field is applied {in the direction parallel to the easy plane}. Since the appearance of the high-field first-order transition stems from the incompatibility between the in-plane quantum magnetic order and the demand of antiparallel alignment along the stacking direction, a similar magnetization process is expected to be obtained for larger spins, e.g., $S=1$ and $S=3/2$. }

{Here, we discussed the case of weak interlayer coupling ($J^\prime =0.025J$) employing the CMF+S based on 2D clusters shown in Fig.~\ref{figclusters}. For larger values of $J^\prime$ ($\gtrsim 0.2 J$), it has been predicted for the Heisenberg model that the umbrella state is stabilized at strong fields~\cite{nikuni-95,marmorini-15}. Therefore, for moderate interlayer couplings, the magnetization process should become more complex due to the competition of the umbrella state and the coplanar states shown in Fig.~\ref{figbranch}. It remains an interesting open problem.}

%------------------------------------------------------------------------------

%\acknowledgment
We acknowledge Hidekazu Tanaka and Yoshitomo Kamiya for useful discussions. This work was supported by KAKENHI Grants from JSPS No. 25800228 (I.D.), No. 25220711 (I.D.), and No. 26800200 (D.Y.).

%\appendix
%\section{}

%Use the \verb|\appendix| command if you need an appendix(es). The \verb|\section| command should follow even though there is no title for the appendix (see above in the source of this file).

%For authors of Invited Review Papers, the \verb|profile| command is prepared for the author(s)' profile.  A simple example is shown below.

\end{document}